\documentclass[sn-nature]{sn-jnl}
\usepackage{graphicx}%
\usepackage{multirow}%
\usepackage{amsmath,amssymb,amsfonts}%
\usepackage{amsthm}%
\usepackage{mathrsfs}%
\usepackage[title]{appendix}%
\usepackage{xcolor}%
\usepackage{textcomp}%
\usepackage{manyfoot}%
\usepackage{booktabs}%
\usepackage{algorithm}%
\usepackage{algorithmicx}%
\usepackage{algpseudocode}%
\usepackage{listings}%

\unnumbered

\newcommand{\kt}{k_\mathrm{B}T}
\newcommand{\ld}{l_\mathrm{D}}
\newcommand{\lb}{l_\mathrm{B}}
\newcommand{\dd}{\mathrm{d}}
\newcommand{\etam}{\eta_\mathrm{m}}
 
\newcommand{\feq}{F_\mathrm{eq}}
\newcommand{\ftot}{F_\mathrm{tot}}
\newcommand{\fnc}{F_\mathrm{nc}}

\makeatletter
\newcommand*{\newbibstartnumber}[1]{%
  \apptocmd{\thebibliography}{%
    \global\c@NAT@ctr #1\relax
    \addtocounter{NAT@ctr}{-1}%
  }{}{}%
}
\makeatother

\begin{document}

\title[Article Title]{Observation of Brownian elastohydrodynamic forces acting on confined soft colloids}


\author[1]{\fnm{Nicolas} \sur{Fares}}\email{nicolas.fares@u-bordeaux.fr}

\author[1,2]{\fnm{Maxime} \sur{Lavaud}}\email{maxime.lavaud@u-bordeaux.fr}
\equalcont{These authors contributed equally to this work.}

\author[1,3]{\fnm{Zaicheng} \sur{Zhang}}\email{zhangzaicheng@buaa.edu.cn}
\equalcont{These authors contributed equally to this work.}

\author[1]{\fnm{Aditya} \sur{Jha}}\email{aditya.jha@u-bordeaux.fr}

\author*[1]{\fnm{Yacine} \sur{Amarouchene}}\email{yacine.amarouchene@u-bordeaux.fr}

\author*[1]{\fnm{Thomas} \sur{Salez}}\email{thomas.salez@cnrs.fr}

\affil[1]{Univ. Bordeaux, CNRS, LOMA, UMR 5798, F-33400, Talence, France}

\affil[2]{Univ. Bordeaux, CNRS, Bordeaux INP, CBMN, UMR 5248, F-33600, Pessac, France}

\affil[3]{School of Physics, Beihang University, 100191 Beijing, China}


\abstract{Confined motions in complex environments are ubiquitous in microbiology. These situations invariably involve the intricate coupling between fluid flow, soft boundaries, surface forces and fluctuations. In the present study, such a coupling is investigated using a novel method combining holographic microscopy and advanced statistical inference. Specifically, the Brownian motion of soft micrometric oil droplets near rigid walls is quantitatively analyzed. All the key statistical observables are reconstructed with high precision, allowing for nanoscale resolution of local mobilities and femtonewton inference of conservative or non-conservative forces. Strikingly, the analysis reveals the existence of a novel, transient, but large, \textit{soft Brownian force}. The latter might be of crucial importance for microbiological and nanophysical transport, target finding or chemical reactions in crowded environments, and hence the whole life machinery.}


\maketitle

\section{Summary}\label{sec:summary}
Confined motions of soft particles are ubiquitous in microbiology. Examples at the micro- and nano-scales include blood cells flowing in vessels, antibody recognition, or confined diffusion of synaptic neurotransmitters. These situations bring about viscous flows coupled to charged and soft confining entities, in presence of thermal fluctuations.
Fluctuation-free scenarios have already proven the emergence of novel softness-induced near-contact forces, but the inclusion of fluctuations -- crucial at microscopic scales -- is yet to be explored.

Aiming at unravelling the link between confined viscous flow, softness and thermal, \textit{i.e.} Brownian, motion, we combine holographic microscopy and statistical inference. This state-of-the art technique allows for a robust, broadband, and non-fluorescent super-resolved characterization of the three-dimensional motion of a single, free, colloid diffusing in a viscous fluid near a charged rigid wall, with nanometric precision. Already in the case of a micrometric rigid polystyrene bead, striking differences compared to bulk Brownian motion are measured and quantified. Namely, the statistics of displacements deviates from Gaussian distributions. Also, femtonewton-like bulk and surface forces, resolved at the fundamental thermal noise limit, which include screened electrostatic repulsion and weight, are extracted from the trajectories.

The novel case of a deformable micro-sphere is even more puzzling. Specifically, we study here the confined Brownian motion of single low-surface-tension viscous oil micro-droplets. 
While, at equilibrium, these soft droplets behave similarly as their rigid counterparts, at short time scales, we observe the emergence of novel transient piconewton-like inertia-less lift forces acting on the droplets, stemming from a thermally-induced visco-capillary coupling. This so-far-overlooked and universal type of forces could drastically impact transient migration strategies, relevant to a myriad of transport situations in microbiology.

\section{Introduction}\label{sec:intro}
An enormous number of simultaneous and overlapping processes ensure the proper functionality of the human body. Nearly all of the underlying mechanisms arise spontaneously through the passive and automated activity of the cells.
Such a general statement raises questions about the fundamental mechanisms leveraged by the human-body machinery at the individual-cell scale which orchestrate greater-scale organization and operation. 
These mechanisms necessarily include the motility of confined microscopic entities, encountered in several situations spanning the micro- and nano-scales: from blood cells flowing in vessels and capillaries~\cite{popel2005microcirculationa,kim2009cellfree,vlahovska2009vesicles}, cell differentiation, to antibody recognition, glia cells' mechano-sensing~\cite{velasco2018infection}, and confined diffusion of synaptic receptors~\cite{dahan2003diffusion,choquet2013dynamic,varela2016targeting} or of neurotransmitters in the synaptic cleft~\cite{armbruster2020effects}. 
The first quest to comprehend these scenarios from a physical standpoint begins with a minimal set of ingredients: viscous flows coupled to charged and soft confining interfaces, in presence of thermal fluctuations. This reasoning yields a twofold state-of-the-art, as presented hereafter.

On the one hand, 
deterministic (\textit{i.e.} fluctuation-free) motion in lubricated geometries~\cite{hamrock2004fundamentals}, where a thin layer of viscous liquid separates an object from a boundary, drastically depends on the softness (\textit{i.e.} ability to deform) of the entities at play~\cite{bureau2023lift,rallabandi2024fluidelastic}. 
This recent area of research, aptly named soft lubrication, revealed the appearance of surprising inertial-like forces and torques, traditionally observed for high-speed and large-scale flows. These include lift forces, Magnus-like effects, or dynamic adhesion, to name a few, which all result from the elastohydrodynamic coupling between the motion-induced viscous flow and the soft confining boundaries~\cite{skotheim2004soft,urzay2007elastohydrodynamic,salez2015elastohydrodynamics,bertin2022softlubrication}. 
Over the past decade, dedicated research has provided decisive experimental evidence for these soft-lubrication effects across a broad range of scales, from the observation of self-sustained lift and reduced friction of a macroscopic cylinder sliding down a soft incline~\cite{saintyves2016selfsustained}, and the observation of a falling bead surfing its own elastic wave over a nearby membrane~\cite{rallabandi2018membraneinduced}, to the direct and quantitative measurement of the soft-lubrication lift force at the nanoscale~\cite{vialar2019compliant,zhang2020direct}. 
Interestingly, such interfacial elastohydrodynamic couplings play a significant role in biological contexts~\cite{brochard2003hydrodynamics}, as demonstrated by the observation of micrometric beads advected in microfluidic channels being drastically repelled from biomimetic hyaluronic acid brushes grafted on rigid walls~\cite{grandchamp2013lift,davies2018elastohydrodynamic}. 
A parameter not to be forgotten resides in the very nature of the interfacial softness. As the entities get softer, or equivalently at smaller scales, capillarity rather than elasticity starts to be the dominant restoring force triggering soft-lubrication couplings~\cite{jha2023capillary}. More precisely, this shift occurs when the sizes involved in the problem become smaller than the elasto-capillary length $\gamma / E$~\cite{roman2010elasto}, typically $1 \, \mathrm{\mu m}$ for a material of Young's modulus $E = 10 \, \mathrm{kPa}$ and surface tension $\gamma = 10 \, \mathrm{mN/m}$, which are typical conditions easily reached both naturally and experimentally. 
Such capillary-induced soft-lubrication forces, coupled to static interactions, were \textit{e.g.} proven to control droplet-droplet interactions~\cite{dagastine2006dynamic}, as well as cell-cell adhesion~\cite{pontani2012biomimetic}. 

On the other hand, 
as the size of the entities decreases, Brownian (\textit{i.e.} thermally-fluctuating or diffusive) motion becomes prominent and is strongly affected by confinement and hydrodynamic interactions~\cite{li2008amplified}. 
Previous research, focused on rigid-confinement scenarios, already pointed out striking differences compared to the idealized bulk picture. 
Faucheux and Libchaber~\cite{faucheux1994confined} demonstrated that diffusion is hindered by confining colloids between rigid walls and opened the way to several related studies~\cite{dufresne2000hydrodynamic,felderhof2005effect,benichou2013geometry,mo2015broadband}. Such research efforts have not only highlighted the complex, anisotropic and space-dependent nature of diffusion in confined spaces~\cite{eral2010anisotropic}, but have also harnessed these features in order to develop novel gentle techniques for nanorheological and surface characterization~\cite{joly2006probing}. Besides, detailed analysis of rigid-confinement scenarios revealed that random walkers also exhibit non-Gaussian statistics of displacements~\cite{guan2014even,matse2017test,alexandre2023nongaussian}, associated with the enhanced occurrence of rare events -- hence with fundamental implications in winners-take-all processes for life.

Ongoing research has yet to explore the \textit{soft Brownian} intersection of the two domains above, namely soft lubrication and confined Brownian motion, which remains theoretically and experimentally challenging to address. Incorporating molecular fluctuations into the non-linear elastohydrodynamic continuum theory is inherently an arduous task. As of yet, theoretical studies remain scarce and have merely focused on the simplified case of point-sized tracers~\cite{bickel2006brownian,marbach2018transport,sarfati2021enhanced}, thereby neglecting any finite-size effects and lubrication couplings. A few alternative numerical strategies have thus emerged~\cite{sheikh2023brownian,shen2023anomalous,zhang2024brownian} and tend to indicate deformability-driven anomalous diffusive behaviors. On the experimental front, evidence for soft Brownian couplings remains elusive~\cite{wang2009hydrodynamic}, which might be attributed to the magnitude of the latter. A coarse hand-waving argument which consists in plugging the thermal root-mean-squared velocity into existing deterministic theories of soft lubrication predicts piconewton-like soft Brownian forces in typical conditions. While this order of magnitude is comparable to the one of other surface and entropic forces -- hence suggesting that it might be biologically relevant in order to revisit and rationalize spontaneous migration and target-finding strategies (\textit{e.g.} for proteins near cell membranes) -- the soft Brownian forces that we propose here remain hypothetical and must then be measured and quantified precisely. Besides such a fundamental challenge, and although more than a century has passed since Brownian motion was first used by Jean Perrin to measure Avogadro's number and prove the atomic hypothesis~\cite{perrin1908agitation}, the intricacies of Brownian motion at complex interfaces could further contribute to open a novel window towards ultra-low force sensing in biological and nanoscale physics~\cite{perrin1908agitation,sainis2007statistics,lavaud2021stochastic}. Fine-tuning such technique remains imminent in the current era, where the demand for miniaturized precise tools is targeted, \textit{e.g.} to probe biologically-relevant elasticity-related phenomena~\cite{kasuba2024mechanical}.

In this article, we address both the above-mentioned fundamental and practical questions \textit{via} our recently-developed non-fluorescent super-resolution microscopy technique based on Mie holography and advanced statistical inference~\cite{lavaud2021stochastic}.
In the subsequent sections, we describe the methods used to generate Brownian oil micro-droplets, track them and analyze their confined motion. From the reconstruction of the whole set of statistical observables, we first robustly recover all the expected equilibrium and dynamical properties with high precision. Using such a calibrated method, we then investigate the transient emergence of piconewton-like non-conservative forces in the combined presence of soft confinement and thermal fluctuations. Mainly, our results seem to indicate the existence of a novel soft Brownian force at short times, that we systematically quantify, characterize and rationalize. 

\section{Experimental situation}\label{sec:method}
\begin{figure}[ht]
    \centering
    \makebox[\textwidth][c]{\includegraphics[scale=0.7]{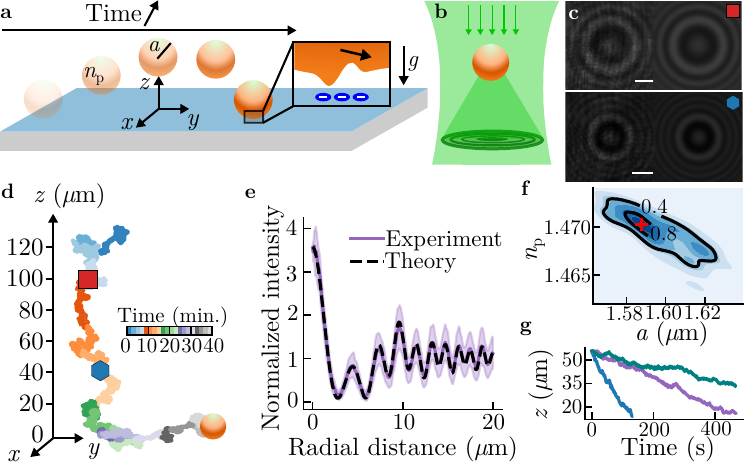}}
    \caption{
    Overview of the experimental method. 
    (\textbf{a}) Schematic of the system: an oil micro-droplet diffuses in a viscous liquid near a rigid glass wall. The inset highlights the surface forces at play: electrostatic repulsion and elastohydrodynamic coupling. 
    (\textbf{b}) Experimental set-up: Mie holography~\cite{lavaud2021stochastic}. 
    (\textbf{c}) Interference patterns (\textit{i.e.} holograms) for two wall-droplet distances $z$, as indicated by the symbols on the top right corners and corresponding ones in (d). Each two-image panel depicts, from left to right, the experimental hologram and the corresponding best-fit Lorenz-Mie one. Scale bars are $5 \, \mathrm{\mu m}$.
    (\textbf{d}) Experimental trajectory of an oil droplet (radius $a = 1.6 \, \mathrm{\mu m}$) diffusing in a water-ethylene-glycol mixture (mass fraction $w = 0.3$). 
    (\textbf{e}) Normalized intensity profiles from the experimental and theoretical holograms for $z = 0.7 \, \mathrm{\mu m}$, as functions of the radial distance to the hologram's center. The light purple area corresponds to the standard deviation. 
    (\textbf{f}) 2D kernel density estimate of the \textit{in-situ} calibration of the radius $a = 1.587 \pm 0.016 \, \mathrm{\mu m}$ and optical index $n_\mathrm{p} = 1.471 \pm 0.002$ of the droplet used in (c,d), measured over $10^4$ points.
    (\textbf{g}) Sedimentation curves used as \textit{in-situ} calibration of medium viscosity $\etam$ and droplet's buoyant density $\Delta\rho$. The three solid lines correspond to the used liquids: ultra-pure DI water (blue, $\etam = 1 \, \mathrm{mPa \cdot s}$, $\Delta\rho = 70 \, \mathrm{kg/m^3}$), 30\%-mass-fraction water-ethylene-glycol mixture (purple, $\etam = 2.1 \, \mathrm{mPa \cdot s}$, $\Delta\rho = 33 \, \mathrm{kg/m^3}$), and 40\%-mass-fraction water-ethylene-glycol mixture (green, $\etam = 2.6 \, \mathrm{mPa \cdot s}$, $\Delta\rho = 18 \, \mathrm{kg/m^3}$).
    }
    \label{fig:setup}
\end{figure}

We generate and employ low-interfacial-tension oil micro-droplets, with radius $a \sim 2 \, \mathrm{\mu m}$ and interfacial tension in water $\gamma = 8 \pm 3 \, \mathrm{mN/m}$, as detailed in Methods and Extended Data Tab.~1. We then study their random trajectories within various viscous immiscible liquids near a rigid glass substrate, as sketched in Fig.~\ref{fig:setup}a. 

Individual micro-droplets in a dilute emulsion are tridimensionally tracked through Mie holography~\cite{lee2007characterizing}, as explained in Methods and depicted in Fig.~\ref{fig:setup}b,c. Tridimensional trajectories $(x(t), \, y(t), \, z(t))$ along time $t$ are reconstructed over broad spatio-temporal ranges, as shown in Fig.~\ref{fig:setup}d. 
This method leads to a spatial tracking resolution of $\sim20~\mathrm{nm}$, thanks to the strong agreement between the experimental and theoretical interference patterns (see Fig.~\ref{fig:setup}c,e).

Moreover, the pattern not only depends on the droplet's tridimensional position, but also on the droplet's radius $a$ and optical index $n_\mathrm{p}$. The two latter parameters are thus measured \textit{in-situ} with precisions of 20 nm and 0.002, respectively, as highlighted in Fig.~\ref{fig:setup}f. 

The trajectories also contain information about the surrounding medium. For every experimental set, the viscosity $\etam$ of the surrounding liquid medium and the relative buoyant density $\Delta\rho$ of the droplet are also measured \textit{in-situ}, by studying the early-stage trajectory (see Methods, Extended Data Fig.~1 and Extended Data Tab.~2), \textit{i.e.} the droplet's sedimentation toward but far from the wall, thus allowing to neglect boundary effects (see Fig.~\ref{fig:setup}d, $z \ge 20 \, \mathrm{\mu m}$). 

The two preliminary steps described above permit to fully characterize each studied droplet, making the method entirely self-calibrated prior to the central statistical study discussed below.

\section{Equilibrium properties}\label{sec:equilibrium}

\subsection{Equilibrium static properties}\label{subsec:static}
\begin{figure}[ht]
    \centering
    \includegraphics[scale=1]{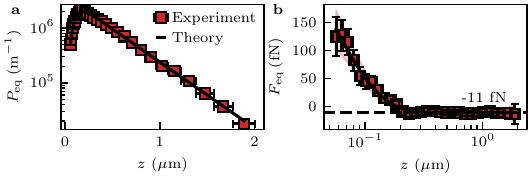}
    \caption{
    Equilibrium static properties for an oil droplet of radius $a = 1.6 \, \mathrm{\mu m}$ diffusing in ultra-pure DI water near a rigid glass wall. The red squares correspond to the experimental data. The black solid lines represent the Boltzmann factor (a) and Eq.~(1) (b), that include electrostatic repulsion and the droplet's buoyant weight. 
    (\textbf{a}) Equilibrium probability distribution function $P_\mathrm{eq}$ as a function of the wall-droplet gap distance $z$. 
    (\textbf{b}) Conservative force $F_\mathrm{eq}$ as a function of $z$. Error bars give a 95\%-confidence interval. The black dashed line indicates the droplet's buoyant weight and the red-colored area the the thermal-noise limit \cite{liu2016subfemtonewton}.
    }
    \label{fig:boltzmann}
\end{figure}

After sedimentation, the droplet reaches equilibrium and steadily diffuses close to the wall. 
It stays confined near the wall as a result of the competition between thermal fluctuations, 
and both the electrostatic repulsion between the negatively-charged surfaces of the droplet and the wall and the droplet's buoyant weight. This translates into Boltzmann's equilibrium probability density function $P_\mathrm{eq}(z)=N\exp[-U_\mathrm{eq}(z)/(k_{\mathrm{B}}T)]$ of the $z$ position (see Fig.~\ref{fig:boltzmann}a), where $N$ is a normalization factor, $U_\mathrm{eq}$ is the potential energy, $k_{\mathrm{B}}$ is Boltzmann's constant and $T$ is the temperature. The total normal conservative force $F_\mathrm{eq}$ acting on the droplet is then defined as:
\begin{equation}
    \label{eq:Feq}
    F_\mathrm{eq} = - \frac{\dd U_\mathrm{eq}}{\dd z} \, ,
\end{equation}
where $U_\mathrm{eq} /(\kt) = Be^{-z/\ld} + z/\lb$. While $B$ quantifies the magnitude of the surface charges, the Debye length $\ld$~\cite{sze2003zetapotential} and the Boltzmann length $\lb = \kt /(\Delta m g)$~\cite{perrin1908agitation} balance thermal energy with the electrostatic repulsion and the buoyant weight $\Delta m g$, respectively, where $\Delta m$ is the buoyant mass and $g$ the acceleration of gravity. Safety checks regarding the measurement of electrostatic properties are given in SI (see Equilibrium Dynamic Properties and Extended Data Figs.~2,3 and Extended Data Tab.~3), in the case of control experiments on rigid polystyrene micro-beads that diffuse in salted water near a rigid glass wall. In particular, the expected values and dependencies on salt concentration for both $\ld$ and $B$ are captured~\cite{behrens2001charge}. 
As shown in Fig.~\ref{fig:boltzmann}b, femtonewton forces are measured and accurately described by Eq.~(1). The error bars are of the same order of magnitude as the theoretical thermal-noise limit~\cite{liu2016subfemtonewton}, meaning that the resolution has reached its ultimate bound for the given system and acquisition time. 

\subsection{Equilibrium dynamic properties}\label{subsec:dynamic}

\begin{figure}[ht]
    \centering
    \makebox[\textwidth][c]{\includegraphics[scale=1]{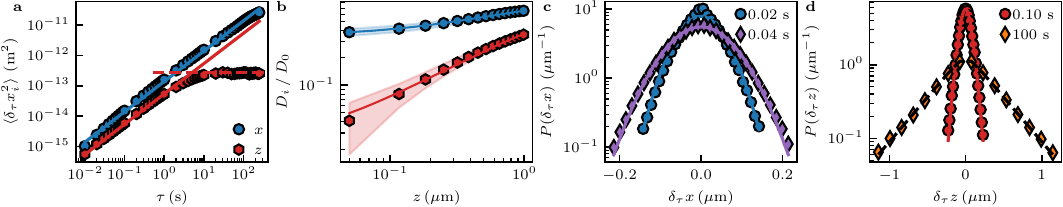}}
    \caption{
    Equilibrium dynamic properties for an oil droplet of radius $a = 1.6 \, \mathrm{\mu m}$ diffusing in ultra-pure DI water near a rigid glass wall.
    (\textbf{a}) Mean-squared displacements $\langle \delta_\tau x_i ^2 \rangle$, as functions of the time increment $\tau$, for the horizontal ($x_i=x$, blue circles) and vertical ($x_i=z$, red hexagons) motions. The solid lines represent the short-time theory, and the dashed line the long-time one \cite{lavaud2021stochastic}.
    (\textbf{b}) Local diffusion coefficients $D_i$ (same color code), normalized by the bulk one $D_0$, as functions of the wall-droplet gap distance $z$. The solid lines represent the theory for the no-slip rigid case~\cite{lavaud2021stochastic}. The colored areas give 95\%-confidence intervals~\cite{vestergaard2015estimation}. 
    (\textbf{c}) Probability density function $P(\delta_\tau x)$ of the transverse displacement $\delta_\tau x$, for the two time increments $\tau$ indicated in legend. The solid lines correspond to the short-time theory \cite{lavaud2021stochastic}. 
    (\textbf{d}) Probability density function $P(\delta_\tau z)$ of the normal displacement $\delta_\tau z$. The red solid line corresponds to the short-time theory, and the black dashed line to the long-time one \cite{lavaud2021stochastic}. Error bars give a 99\%-confidence interval.}
    \label{fig:displacement}
\end{figure}

A complete description of equilibrium invariably includes dynamic observables. For clarity purposes, the theoretical framework is provided in Supplementary Information (SI). The canonical dynamic variable is the mean-squared displacement (MSD) $\langle \delta_\tau x_i ^ 2 \rangle$ ($x_i=x,z$), shown in Fig.~\ref{fig:displacement}a, where $\delta_\tau x_i$ denotes the displacement of the coordinate $x_i$ over the time increment $\tau$, and where $\langle\cdot\rangle$ indicates a temporal average. The plot shows that, as in the bulk and rigid-confinement~\cite{lavaud2021stochastic} cases, the MSDs are linear in time, at short times. The process is thus Brownian (or Fickian). The saturation observed in the vertical direction at longer times simply results from the effective trapping due to gravity. 
Besides, as in the rigid-confinement case, the droplet's motion is anisotropic, resulting in a smaller average diffusion coefficient in the vertical direction as compared to the horizontal one. By binning the data in $z$ and using an advanced inference method (see Inference of Local Diffusion Coefficient Profiles in SI)~\cite{frishman2020learning}, the local diffusion coefficients can also be measured with a nanometric spatial resolution (see Fig.~\ref{fig:displacement}b). In all cases, the measurements agree with the theoretical predictions for rigid no-slip confinement~\cite{faxen1923bewegung,brenner1961slow}. Furthermore, the probability density functions of the transverse and normal displacements, $P(\delta_\tau x)$ and $P(\delta_\tau z)$ respectively, are measured at short and long times (see Fig.~\ref{fig:displacement}c,d). Again, the experimental data matches the theory established for rigid no-slip confinement (see Equilibrium Dynamic Properties in SI). 

In summary, the experiments on soft micro-droplets are accurately captured at equilibrium by the rigid-colloid theoretical framework, as it is the case for independent control experiments with rigid polystyrene micro-beads (see Extended Data Fig.~2).
Hence, at equilibrium, a soft micro-droplet behaves as its rigid counterpart. However, this statement breaks down at short time scales, as discussed hereafter.

\section{Main result}

\begin{figure}[ht]
    \centering
    \makebox[\textwidth][c]{\includegraphics[scale=1]{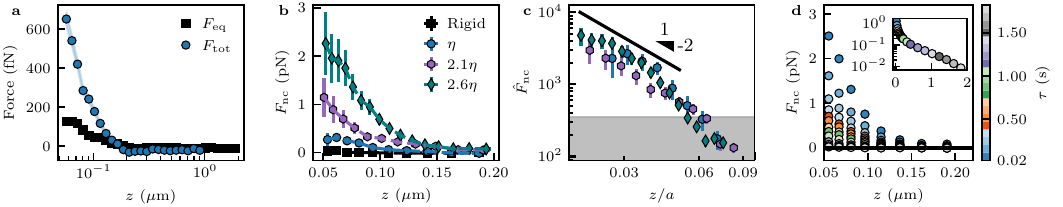}}
    \caption{
    Transient inertia-less visco-capillary lift force acting on micro-droplets. 
    (\textbf{a}) Equilibrium force $F_\mathrm{eq}$ (black squares, see Eq.~(1)), and total force $F_\mathrm{tot}$ (blue circles, see Eq.~(2)) in excess to the Stokes-like drag, as functions of the wall-droplet gap distance $z$.
    (\textbf{b}) Non-conservative force $F_\mathrm{nc} = F_\mathrm{tot} - F_\mathrm{eq}$ in excess to the Stokes-like drag, as a function of $z$. The black squares correspond to the case of rigid polystyrene colloids. The other colored symbols correspond to soft oil droplets in different viscous liquids.
    The outer-liquids' viscosities $\etam$ relative to the one of water $\eta$ are indicated in legend. 
    (\textbf{c}) Normalized non-conservative force $\hat{F}_\mathrm{nc} = F_\mathrm{nc} \, (\gamma \rho_\mathrm{p} a^2) / (\etam^2\kt)$ as a function of the dimensionless gap $z/a$. Symbols correspond to the ones in (B). The black solid line corresponds to Eq.~(4), and the gray area to the normalized thermal-noise limit (see Fig.~\ref{fig:boltzmann}b).
    (\textbf{d}) Non-conservative force $\fnc$ as a function of $z$, for different time increments $\tau$, in the case of a 2-$\mathrm{\mu m}$ droplet diffusing in a 40-\% water-ethylene-glycol mixture. 
    The inset shows the average over $z$ of $\fnc$ (in pN), as a function of $\tau$ (in s). 
    Error bars are removed for clarity.
    }
    \label{fig:Fnc}
\end{figure}

In addition to the conservative force obtained above from the equilibrium distribution in vertical position, one can measure from the vertical drifts the total vertical force $F_\mathrm{tot}$ (including conservative and possible novel non-conservative forces) in excess to the no-slip rigid-confinement Stokes-like drag, as:
\begin{equation}
    \label{eq:Ftot}
    F_\mathrm{tot} (z) = \frac{\kt}{D_z} \, \left( \frac{\langle \delta_\tau z \rangle_z}{\tau} - \frac{\dd D_z}{\dd z} \right) \, ,
\end{equation}
where the terms on the right-hand side correspond to the Stokes drag and the spurious drift (see Measurement of Non-Conservative Forces in SI), and where $\langle\cdot\rangle_z$ denotes the average over a vertical bin. The measured $\ftot$ contains the conservative and non-conservative contributions, except for Stokes drag by construction. 
This total force $\ftot$ perfectly matches the previously-mentioned conservative force $F_\mathrm{eq}$ in the case of rigid polystyrene spherical colloids (see Extended Data Fig.~4), indicating that no novel non-conservative force $\fnc = \ftot - \feq$ acts on rigid colloids, as expected. 
Conversely, $\fnc$ appears to be non-zero, positive and large (\textit{i.e.} in the piconewton range) as compared to the conservative force, in the case of confined soft droplets at early time scales (see Fig.~\ref{fig:Fnc}a). This observation seems to indicate the spontaneous emergence of some out-of-equilibrium vertical force. Furthermore, this force increases with the outer-medium viscosity $\etam$ (see Fig.~\ref{fig:Fnc}b) in a super-linear, quadratic-like manner (see Fig.~\ref{fig:Fnc}c), possibly suggesting a non-linear hydrodynamic origin. Besides, it decays as the droplet moves at larger distances from the wall, highlighting some interfacial or confinement-induced origin. 
Finally, this force appears to be transient, as evidenced by the temporal dependency of $\fnc$ shown in Fig.~\ref{fig:Fnc}d. Indeed, as the time increment $\tau$ used to evaluate $\fnc$ increases, $\fnc$ decreases and converges toward 0 over a few seconds (see inset in Fig.~\ref{fig:Fnc}d). Interestingly, this time scale is comparable to the time  $\tau_{\mathrm{eq}}$ required to reach equilibrium in the system, \textit{i.e.} to observe the transition from free diffusion to effective trapping on the MSD of the vertical motion (see Fig.~3A). The inset in Fig.~\ref{fig:Fnc}d further suggests that such a transient non-conservative force may vanish toward equilibrium in an exponential-like manner. 

\section{Discussion}\label{sec:discussion}
In the reminder of the article, we argue that a relevant candidate to explain all these quantitative and systematic observations is the soft Brownian force proposed above. First, fluid inertia can be disregarded, as the typical inertial time $\tau_\mathrm{in} \sim \rho_\mathrm{m} a^2 / \etam \sim 10 \, \mathrm{\mu s}$, where $\rho_\mathrm{m}$ denotes the outter-medium density, is negligible compared to the minimal time increment $\tau = 10 \, \mathrm{ms}$ used to measure the drifts $\delta_\tau z$. 
Hence, the outer-fluid flow induced by the droplet motion is governed by the steady incompressible Stokes equation which, in the near-contact lubricated regime where $z \ll a$, reads $\partial_r \, p = \etam \, \partial_{zz} \, v$, where $p$ and $v$ respectively denote the hydrodynamic pressure and velocity fields in the fluid gap between the droplet and the wall, and $r$ is the radial horizontal coordinate. Moreover, the droplet is much more viscous than the outer medium (viscosity ratio greater than 400), so that we can neglect fluid flow inside the droplet. Carrying on with the modeling (see details in SI), we focus on the horizontal motion of the droplet with a typical velocity $V_0$. 
Under all these assumptions, and in the limit of small interfacial deformations $\delta$ (\textit{i.e.} $\delta \ll z$), the pressure $p$ scales as:
\begin{equation}
    \frac{p}{l} \sim \etam \frac{V_0}{(z+\delta)^2} \simeq \frac{\etam V_0}{z^2} \, \left( 1 - 2\frac{\delta}{z} \right) \, ,
\end{equation}
where $l = \sqrt{2az}$ is the hydrodynamic radius, \textit{i.e.} the horizontal extent of the lubrication pressure field \cite{skotheim2004soft}.
The first and second terms in the right-hand side of Eq.~(3) correspond to the zeroth- and first-order pressure contributions of the perturbation expansion in the deformation. 
The former, akin to the excess pressure field generated from the translational lubricated motion of a rigid bead near a rigid wall with no slippage~\cite{faxen1923bewegung}, does not generate any net vertical force by integration because it is an antisymmetric function due to both the time-reversal symmetry of the Stokes equation and the even contact geometry. However, it deforms the droplet according to Laplace's law $p \sim  \gamma \, \delta / l^2$.

As a consequence of the flow-induced droplet's deformation, the above-mentioned fore-aft contact symmetry is broken, and the first-order pressure then leads to a non-zero vertical force. After combining Laplace's law with Eq.~(3), and integrating the first-order pressure over the area $\pi l^2$, this vertical force reads $\sim \etam^2V_0^2a^3/(\gamma z^2)$. Further using the thermal velocity $\sqrt{3\kt / (2\pi a^3 \rho_\mathrm{p})}$ as an estimate for the horizontal velocity scale $V_0$ in the fluctuating case, and adding an ad-hoc exponential cut-off factor towards equilibrium, one gets the following scaling suggestion for the soft-Brownian force:
\begin{equation}
    F_{\mathrm{nc}} \sim \frac{\etam^2\kt}{\gamma\rho_\mathrm{p}a^2} \, \left( \frac{a}{z} \right) ^2\exp\left(-\frac{\tau}{\tau_{\mathrm{eq}}}\right) \, .
\end{equation}

As observed in Fig.~\ref{fig:Fnc}c, Eq.~(4) allows to collapse all the experimental data into a single master curve for a given time lapse $\tau$. Furthermore, the correct $\sim (z/a)^{-2}$ power law is recovered above the thermal-noise limit of the method, with a multiplicative factor similar in  order of magnitude as the one estimated theoretically (see Soft Lubrication Model in SI). All together, these elements seem to indicate that the transient but large non-conservative force measured in our study does indeed emerge from an intricate coupling between softness, flow confinement and thermal fluctuations. A final remark should nevertheless be made on the fluid's inertia contribution to the droplet's dynamics and resulting force. The total force measurement is performed by capturing stroboscopic glimpses of the droplet's trajectory acquired at a frequency $1/\tau$, which is smaller than $1/\tau_\mathrm{in}$. While the measured force seems to be reminiscent of inertial dynamics, the actual droplet's mass to consider in the velocity scale is not trivial. Indeed, fluid inertia does classically contribute to the effective mass of immersed objects in the bulk, but i) the added mass becomes space-dependent in rigid confinement~\cite{zhang2023unsteady}, and ii) we might even anticipate novel softness-induced corrections to the added mass. Further experimental an  theoretical investigations of fluctuating elastohydrodynamics at small time scales are thus needed in the future in order to address such a matter. 

\section{Conclusion}\label{sec:ccl}

Using a state-of-the-art, broadband, contactless and non-fluorescent method based on optical holography and statistical inference, we reported the first observation of a transient but large spontaneous colloidal \textit{soft Brownian force}, solely triggered by the coupling between thermal fluctuations, confined flow and soft boundaries. This result is believed to be significant for nanoscience, as such a so-far-overlooked effect might strongly contribute to short-term microscopic transport, target finding and chemical reactions in complex bounded environments -- that are all ubiquitous features in microbiological and nanophysical settings. 

\backmatter

\bmhead{Supplementary information}

Supplementary information is available for this paper. A short experimental video is also shared on the open-access repository:
\newline
https://github.com/EMetBrown-Lab/Brownian-motion-near-soft-interfaces. 

\bmhead{Data availability}

All data and codes used in this study are available on the open-access repository: 
\newline
https://github.com/EMetBrown-Lab/Brownian-motion-near-soft-interfaces. 

\bmhead{Acknowledgements}

The authors thank Nicolas Bain, Kari Dalnoki-Veress, Carlos Drummond, Aurélie Hourlier-Fargette, Frédéric Restagno, Jean-Baptiste Salmon, Robert Style, Emilie Verneuil, and Kaili Xie for useful insights on materials. They thank Maxence Arutkin, Vincent Bertin, David Dean, James Forrest, Juliette Lacherez, Yann Louyer, Joshua McGraw, Elodie Millan, Elie Raphaël, Raphaël Sarfati, Pierre Soulard, and Yilin Ye for interesting discussions. They also thank Josiane Parzych for financial project management, as well as Romain Houques and Anne Tempel for lab-space installation. 

The authors acknowledge financial support from the European Union through the European Research Council under EMetBrown (ERC-CoG-101039103) grant. Views and opinions expressed are however those of the authors only and do not necessarily reflect those of the European Union or the European Research Council. Neither the European Union nor the granting authority can be held responsible for them. The authors also acknowledge financial support from the Agence Nationale de la Recherche under EMetBrown (ANR-21-ERCC-0010-01), Softer (ANR21-CE06-0029) and Fricolas (ANR-21-CE06-0039), as well as from the Interdisciplinary and Exploratory Research program under MISTIC grant at University of Bordeaux, France. They also acknowledge the support from the LIGHT S\&T Graduate Program (PIA3 Investment for the Future Program, ANR-17EURE-0027). Finally, they thank the RRI ``Frontiers of Life", which received financial support from the French government in the framework of the University of Bordeaux's France 2030 program, as well as the Soft Matter Collaborative Research Unit, Frontier Research Center for Advanced Material and Life Science, Faculty of Advanced Life Science at Hokkaido University, Sapporo, Japan.

\bmhead{Author contributions}

N.F., M.L. and Z.Z. performed the experiments. N.F. performed the data analysis and visualization. M.L. developed the tracking and analysis algorithms. Z.Z. developed the droplet nebulization technique. A.J. performed the theoretical modeling. Y.A. and T.S. designed and supervised the project. T.S. obtained funding. N.F., Y.A. and T.S. worked on the interpretation and validation of the results. N.F. wrote the initial draft of the manuscript. M.L., A.J. and T.S. edited the manuscript. Z.Z. and Y.A. provided additional comments and ideas on the manuscript. 

\bmhead{Competing interests}

The authors declare no competing interests.



\bibliographystyle{sn-nature}

\section{Methods}\label{sec:methods}

\subsection{Materials}
The oil micro-droplets are made by mixing $200 \, \mathrm{\mu L}$ of oil (AR1000, purchased from Wacker\copyright) and $10 \, \mathrm{mL}$ of the working viscous liquid. The mixture is then put in an ultrasonic bath for ten minutes, and diluted. This procedure leads to oil droplets of radii ranging from $1 \, \mathrm{\mu m}$ to $5 \, \mathrm{\mu m}$, with high reproducibility~\cite{crisp1987effect,heidenkummer1991emulsification}. Their interfacial tension $\gamma = 8 \pm 3 \, \mathrm{mN/m}$ in water is measured by the pendent-drop method~\cite{daerr2016pendent_drop}. Results for different experimental conditions are presented in Extended Data Tab.~1.
\newline
\indent Control experiments are conducted with rigid polystyrene micro-spheres purchased from Polysciences\copyright. From now on, if the distinction is not necessary, the rigid polystyrene micro-spheres or the soft oil micro-droplets will be referred to as colloids. 
\newline 
\indent Chambers containing the diffusing colloids are made of one glass slide and one glass cover-slip (purchased from Academy\copyright), and sealed thanks to melted Parafilm\copyright.
\newline 
\indent The working viscous liquids refer to mixtures of pure DI water (type 1, MilliQ\copyright device) with ethylene glycol (anhydrous, 99.8\%, purchased from Sigma Aldrich\copyright). Properties of the mixtures are described in the caption of Fig.~1g of the main text and Extended Data Tab.~2. 

\subsection{Tridimensional tracking of single colloids}\label{subsec:tracking}
Individual micrometric colloids in a dilute suspension are tridimensionally tracked through Mie holography \cite{lee2007characterizing,bohren2008absorption}. A coherent plane wave (wavelength $532 \,\mathrm{nm}$) illuminates an isolated colloid and is thus scattered. The scattered light interferes with the incident one, leading to a pattern called hologram (see Figs.~\ref{fig:setup}b,c). 
Holograms are recorded \textit{via} an x100 oil-immersed Olympus objective (numerical aperture: 1.45) for one hour at a frame rate of 100~Hz, and fitted according to the Lorenz-Mie theory thanks to the Pylorenzmie software suite~\cite{altman2022holographic}, allowing us to reconstruct the tridimensional trajectories $(x(t), \, y(t), \, z(t))$ along time $t$, over broad spatio-temporal ranges, as shown in Fig.~\ref{fig:setup}d. Experimental spacial illumination defects and variations are corrected by normalizing the image by the background image. The latter is obtained by computing the median image of the movies. 
Also, parameters which are independent of the colloid but affect the holograms are measured separately and fixed. The liquid's refractive index $n_\mathrm{m}$ is measured thanks to a refractometer (DR-M2/1550 ATAGO\copyright) and the results are indicated in Extended Data Tab.~2. The illumination comes from a collimated laser diode (CPS532 from Thorlabs\copyright). 

This method leads to a spatial tracking resolution of $\sim20~\mathrm{nm}$, thanks to the strong agreement between the experimental and theoretical holograms (see Figs.~\ref{fig:setup}c,e).

\subsection{\textit{In-situ} calibration}
Two preliminary steps permit to fully characterize each studied droplet, making the method entirely self-calibrated prior to the central statistical study discussed in the main text.

First, the fitting not only depends on the droplet's tridimensional position, but also on the droplet's radius $a$ and optical index $n_\mathrm{p}$. 
The two latter parameters are calibrated \textit{in-situ} by fitting ten thousand holograms and finding the most probable $a$-$n_\mathrm{p}$ combination (see Fig.~\ref{fig:setup}f), with precisions of 20 nm and 0.002, respectively.
The usefulness and precision of this self-calibration is highlighted through experiments on both polystyrene beads and oil droplets. 
The control experiments are conducted with polystyrene beads of nominal radii of $1.55 \pm 0.03 \, \mathrm{\mu m}$ (13 experiments) and $2.965 \pm 0.15 \, \mathrm{\mu m}$ (13 experiments). Measured radii are $1.50 \pm 0.02 \, \mathrm{\mu m}$ and $2.98 \pm 0.03 \, \mathrm{\mu m}$, respectively, where all uncertainties correspond to a 99-\% confidence interval. So, the nominal and measured values are in agreement up to experimental errors \textit{i.e.} $20 \, \mathrm{nm}$. As for $n_\mathrm{p}$, these experiments give $n_\mathrm{p} = 1.589 \pm 0.005$, again in agreement with the expected value.
Now, over 32 distinct experiments on droplets of unknown size, one gets $n_\mathrm{p} = 1.461 \pm 0.003$ (average $\pm$ standard deviation), which is in agreement with the nominal value up to 0.1 \% (relative error).
\smallskip
\newline
\indent Second, independently of the parameters that affect the optical Lorenz-Mie theory, other parameters can be measured \textit{in-situ} thanks to the available trajectories. Specifically, before getting confined, a droplet falls toward the wall. 
Its sedimentation speed $v_\mathrm{s}$, arising from the balance of viscous dissipation and gravity acceleration $g$, reads:
\begin{equation}
    \label{eq:vs}
    v_\mathrm{s} = \frac{2}{9} g \, a^2 \frac{\Delta\rho}{\etam} \, . 
\end{equation}
Thus, measuring $v_\mathrm{s}$ gives the buoyant density $\Delta\rho$, \textit{i.e.} the mismatch between the medium and the falling droplet, as long as the liquid's bulk viscosity $\etam$ is known. 
Both $v_\mathrm{s}$ and $\etam$ are measured \textit{in-situ} from the sedimentation. 
The former, $v_\mathrm{s}$, is computed through the vertical drift $\langle \delta_\tau z \rangle$ (see Extended Data Fig.~1a), where $\delta_\tau z = z(t+\tau) - z(t)$ denotes a vertical displacement at a time $t$ and over a time delay $\tau$, and $\langle\cdot\rangle$ the average over time. 
The latter, $\etam$, is computed through the second-order moment $\langle \delta_\tau x ^2 \rangle$ of the horizontal displacement $\delta_\tau x$ (see Extended Data Fig.~1b). 
For a falling colloid, $\langle \delta_\tau z \rangle = v_\mathrm{s}\tau$, and $\langle \delta_\tau x ^2 \rangle = 2D_0\tau$, where $D_0 = \kt / (6\pi\etam a)$ denotes the bulk diffusion coefficient and $\kt$ the thermal energy.
Note that the cut-off of the full trajectory is chosen so that the colloid remains at a distance from the wall ten time greater than $a$, in order to ensure measuring bulk properties. 

Finally, those \textit{in-situ} micro-scale measurements are compared to \textit{ex-situ} macro-scale rheological ones. Each liquid is sheared in a cone-plate geometry of a rheometer (HR20 from TA Instruments\copyright), 10 times, with shear rates between $100 \, \mathrm{s^{-1}}$ and $1000 \, \mathrm{s^{-1}}$. 
The \textit{in-situ} and \textit{ex-situ} measurements, summarized in Extended Data Tab.~2, are consistent with each other up to 10~\%, as are the resulting density mismatches $\Delta\rho$ and the expected ones from literature~\cite{volk2018density,fogg1955densities}, up to the experimental error. 

\subsection{Soft-Lubrication Model}\label{subsec:softlub}
The main points precising the scaling discussed in the main text for the soft Brownian force are given hereafter. Specifically, the deterministic expression for the lift force acting on an immersed droplet moving horizontally near a rigid wall is first derived, and then adapted to the fluctuating case through a simple argument on the velocity scale. The complete derivation is provided in SI.

We specifically focus on the case where the viscosity of the droplet, denoted by $\eta_\mathrm{p}$, is much larger than the viscosity of the liquid in the outer medium. Along with a no-slip criterion at the interface of the droplet, this corresponds to the droplet translating with the velocity of the center of mass of the droplet like a rigid bead. The velocity of the center of mass is denoted by $V_0$ and is aligned with the $x$-axis. To highlight the influence of the boundary on the droplet's motion, we focus on the lubrication regime where the gap distance $z \ll a$. Given that the dimensions of the droplet are on the order of a few microns and that it translates with a small velocity of the order of the thermal velocity, the Reynolds number $Re = \rho_\mathrm{m} V_0 a / \etam$ is much smaller than unity, allowing us to ignore inertial effects in the fluid. 

The fluid flow is described by Stokes' equations and the incompressibility condition, associated with no-slip boundary conditions at both the droplet surface and the rigid boundary. 
The interface of the droplet deforms based on the Young-Laplace's law allowing us to define compliance $\kappa = \textrm{Ca}/\epsilon^3$ of the interface deformation of the droplet, where $\textrm{Ca} = \etam V_0/\gamma$ is the capillary number which quantifies the strength of viscous dissipation in comparison to surface tension, and $\epsilon = z/l$ denotes the standard lubrication parameter. For small compliance, the lubrication pressure can be calculated perturbatively with the leading order pressure field corresponding to that near a rigid boundary~\cite{o1964slow}. Since the nature of the interface deformation can be calculated, the perturbative correction to the pressure field dependant on the interface deformation can then be calculated as well. 
This $O(\kappa)$ pressure field leads to the soft-lubrication vertical force $F_+$. After integration:
\begin{equation}
    \label{eq:force_liquid}
    F_+ \simeq 16\pi \left(\frac{\etam^2 V_0^2 a^3}{\gamma z^2}\right)\int_0^{\infty} RP_{10}\textrm{d}R = \frac{6\pi}{25}\frac{\etam^2 V_0^2 a^3}{\gamma z^2}~. 
\end{equation}
Using the squared thermal velocity $3\kt / (2\pi a^3 \rho_\mathrm{p})$ (for two horizontal degrees of freedom) as an estimate of $V_0^2$ in the fluctuating case, plugging it into Eq.~(\ref{eq:force_liquid}), and
adding an ad-hoc exponential cut-off factor to ensure long-term equilibrium, one finally gets a minimal Ansatz for the soft-Brownian force:  
\begin{equation}
    F_{\mathrm{nc}} \approx 0.36\,  \frac{\etam^2\kt}{\gamma\rho_\mathrm{p}a^2} \, \left( \frac{a}{z} \right) ^2\exp\left(-\frac{\tau}{\tau_{\mathrm{eq}}}\right) \, ,
\end{equation}
where $\tau_{\mathrm{eq}}$ is the equilibration time.

\newbibstartnumber{58}

\section{Extended Data}

\backmatter

\bmhead{Extended Data Fig.~1} 
Measurement of the liquid's viscosity $\etam$ and the colloid's buoyant density $\Delta\rho$. Observables computed from the trajectory of an oil droplet ($a = 1.6 \, \mathrm{\mu m}$) falling toward a rigid glass wall in a water-ethylene-glycol mixture (mass fraction 0.3). 
(\textbf{a}) Vertical drift $\langle \delta_\tau z \rangle$ and (\textbf{b}) parallel MSD $\langle \delta_\tau x^2 \rangle$, as a function of the time increment $\tau$. The lines correspond to the best fits to the theories described in Methods (\textit{In-Situ} Calibrations).

\bmhead{Extended Data Fig.~2} 
Summary plot of the observables used in the simultaneous-fitting procedure, in the case of a control experiment: a rigid polystyrene bead of radius $a = 1.48 \pm 0.01 \, \mathrm{\mu m}$, diffusing in salted water ($\ld = 16 \pm 3 \, \mathrm{nm}$) on top of a rigid glass surface. The symbols correspond to experimental data, and the lines to the best fits to the corresponding  theories. 
(\textbf{a}) and (\textbf{b}) See Equilibrium Static Observables in SI. 
(\textbf{c}), (\textbf{d}), (\textbf{e}), and (\textbf{f}) See Equilibrium Dynamic Properties in SI. 
(\textbf{f}) The colored areas symbolize the theoretical error on the local diffusion coefficients $D_i$~\cite{vestergaard2015estimation}.

\bmhead{Extended Data Fig.~3}
Measurement of electrostatic properties.
(\textbf{a}) Debye length $\ld$ as a function of salt concentration [NaCl], in the case of a rigid polystyrene bead in salted water (black squares). The solid line corresponds to the theoretical expectation (see Equilibrium Static Properties in SI), with no free parameter. 
(\textbf{b}) Magnitude $B$ of the electrostatic repulsion between the wall and the polystyrene bead of radius $a = 1.5 \, \mathrm{\mu m}$ (black squares), as a function of the salt concentration [NaCl]. 
(\textbf{c}) Magnitude $B$ of the electrostatic repulsion between the wall and either a polystyrene bead (black squares) or an oil droplet (blue circles), as a function of the colloid's radius $a$. The dashed line indicates the expected linear dependency~\cite{behrens2001charge}, with the surface charge of the colloid as a free parameter. 

\bmhead{Extended Data Fig.~4}
Proof that there is no non-conservative force in the case of the control experiments. Different forces measured as functions of the gap distance $z$ between a rigid polystyrene bead (radius $a = 1.48 \pm 0.01 \, \mathrm{\mu m}$) diffusing in salted water ($\ld = 16 \pm 3 \, \mathrm{nm}$) and a rigid glass wall. The legend indicates the experimental equilibrium conservative force $\feq$ (see Eq.~\ref{eq:Feq}), the experimental total force (in excess to Stokes-like drag) $\ftot$ (see Eq.~\ref{eq:Ftot}), and the theoretical thermal-noise limit $\Delta F$ (see~\cite{liu2016subfemtonewton} and SI).

\bmhead{Extended Data Fig.~5}
Schematic of the near-contact lubrication region of the system. The oil droplet is characterized by its mass density $\rho_{\mathrm{p}}$ and viscosity $\eta_{\mathrm{p}}$. The outer water-based solution is characterized by its mass density $\rho_{\mathrm{m}}$ and viscosity $\eta_{\mathrm{m}}$. The underformed profile of the thin lubrication (water-based) film is noted $h$ and the deformation of the capillary interface is noted $\delta$.

\bmhead{Extended Data Tab.~1}
Interfacial tension $\gamma$ measured by the pendent-drop method, for several types of droplet in several external fluids.

\bmhead{Extended Data Tab.~2}
Measured refractive index $n_\mathrm{m}$, liquid viscosity $\etam$ (in $\mathrm{mPa\cdot s}$) and buoyant density $\Delta\rho$ (in $\mathrm{kg/m^3}$) of an oil micro-droplet, for the mixtures used in the study. The percentage indicates the weight fraction of ethylene glycol (EG). The terms \textit{ex-situ} and \textit{in-situ} refer to the rheometer and holography measurements, respectively.

\bmhead{Extended Data Tab.~3}
Results of the fitting procedure for control experiments, \textit{i.e.} with rigid polystyrene bead diffusing atop a nearby rigid glass surface, in salted water. All the values are indicated in nanometers. Each line corresponds to a 1-hour trajectory at 100 frames per second, \textit{i.e.} 360 000 points.

\end{document}